\documentclass[12pt]{article}
\begin{document}
\baselineskip=16pt

\title
{\begin{Huge}
Are Black Holes Elementary Particles ?\\
\end{Huge}
\vspace{1in} }
\author{Yuan K. Ha\\ Department of Physics, Temple University\\
 Philadelphia, Pennsylvania 19122 U.S.A. \\
 yuanha@temple.edu \\    \vspace{.1in}  }
\date{March 1, 2009}
\maketitle
\vspace{.5in}
\begin{abstract}
\noindent
Quantum black holes are the smallest and heaviest conceivable elementary particles. They have a
microscopic size but a macroscopic mass. Several fundamental types have been constructed with
some remarkable properties. Quantum black holes in the neighborhood of the Galaxy could resolve
the paradox of ultra-high energy cosmic rays detected in Earth's atmosphere. They may also play
a role as dark matter in cosmology.\\
\end{abstract}

\newpage
\noindent
{\bf 1 \hspace{.1in} Introduction}\\

\vspace{.2in}
\noindent
In physics, there are two theoretical lengths for an object: the classical size and the quantum size. The classical size of an object is given by its classical radius in a classical theory, whereas the quantum size is given by the Compton wavelength in quantum mechanics.  For example, the classical radius of the electron is given by
\begin{equation}
r = \frac{e^2}{mc^2} = 2.82 \times 10^{-13} \hspace{.1in} {\rm cm,}
\end{equation}                                                                          
where $e$ is the electron's charge in cgs units, $m$ is its mass and $c$  is the speed of light. The Compton wavelength of the electron is given by
\begin{equation}
\lambda = \frac{\hbar}{mc} = 2.42 \times 10^{-10} \hspace{.1in} {\rm cm,}                                                
\end{equation}
where $\hbar$ is Planck's constant. In general, if the classical radius of an object is larger than its Compton wavelength, then a classical description is sufficient.  On the other hand, if the Compton wavelength of the object is larger than its classical size, then a quantum description is necessary. The electron is a quantum particle. For a black hole, the Schwarzschild radius
\begin{equation}
R_{S} = \frac{2GM}{c^{2}}
\end{equation}
is proportional to the mass $M$, with $G$ being the gravitational constant, but the Compton wavelength
\begin{equation}
\lambda = \frac{\hbar}{Mc}
\end{equation}                                                                                                                         
is proportional to the inverse mass. When the two lengths become equal, a quantum black hole is formed. This occurs at the Planck mass given by
\begin{equation}
M_{Pl} = \sqrt{ \frac{\hbar c}{G} }= 2.2 \times 10^{-5} \hspace{.1in} {\rm gm.}
\end{equation}
The corresponding length is the Planck length 
\begin{equation}
l_{Pl} = \sqrt{ \frac{\hbar G}{c^{3}} } = 1.6 \times 10^{-33} \hspace{.1in} {\rm cm.}
\end{equation}

\newpage
\noindent
{\bf 2 \hspace{.1in} The Nature of Gravity}\\

\vspace{.2in}
\noindent
Quantum black holes are the smallest and heaviest conceivable elementary particles [1].  They have a microscopic size but a macroscopic mass. They exist at the boundary between classical and quantum regions. They obey the Laws of Thermodynamics and they decay into elementary particles. As semi-classical objects, they may subject to the rules of quantum mechanics but not necessarily to the rules of quantum field theory. At present, there is a total lack of evidence of any quantum nature of gravity, despite intensive efforts to develop a quantum field theory of gravity. Is it possible that gravity is an intrinsically classical theory? In general relativity, spacetime is a macroscopic concept. Is Einstein's equation similar in nature to Navier-Stokes equation in fluid mechanics as a macroscopic theory? We notice a similar situation in nuclear physics. It is well known that the energy levels in many nuclei are quantized and they obey the rules of quantum mechanics, but nuclear force is not a fundamental force. When a quantum field theory of nuclei was attempted the resulting theory became inconsistent. The underlying theory for nuclei is now Quantum Chromodynamics constructed in terms of quarks and gluons.\\

\vspace{.25in}
All quantum field theories of gravity require the graviton, a hypothetical spin-2 massless particle. The existence of the graviton in nature remains to be seen. At best it propagates in an {\em a priori} flat background spacetime. This is because the gravitational wave equation,
\begin{equation}
\left( \frac{1}{c^{2}} \frac{\partial^{2}}{\partial t^{2}} - \nabla^{2} \right) h_{\mu\nu} = 0,
\end{equation}                                                                                    
from which the graviton idea is developed, is inherently a {\em weak field} approximation in general relativity. In contrast, the same wave equation applied to the electromagnetic fields and leading to the concept of the photon is an {\em exact} derivation of Maxwell's equations. In the above equation, the gravitational field tensor $h_{\mu\nu}$ is obtained from the metric tensor $g_{\mu\nu}$ by setting
\begin{equation}
g_{\mu\nu} = \eta_{\mu\nu} + h_{\mu\nu},
\end{equation}
where $\mid h_{\mu\nu} \mid \ll 1$, and $\eta_{\mu\nu}$  is the Minkowskian spacetime metric with signature $(1,-1,-1,-1)$. However, it is physically impossible to detect a single graviton not only because of its very low frequency $\omega$ and hence its low energy $\hbar \omega$, but upon analysis of its reasonable probability of capture by a detector, it is found that the detector size $R$ has to be less than the Schwarzschild radius $R_{S}$ of the detector itself [2].  This makes it almost impossible to verify a quantum field theory of gravity based on the graviton. Besides, what is the graviton when gravity is strong? A quantum theory of gravity based on a flat background spacetime is contradictory to the dynamical nature of spacetime in general relativity. If one believes that general relativity is a macroscopic theory similar to Navier-Stokes theory, then there is the possibility that gravity is not a fundamental interaction, in which case quantizing general relativity is similar to quantizing the strong nuclear force, leading inevitably to an inconsistent theory. The present impasse in quantizing general relativity after decades of persistent but futile efforts may be an indication that gravity is intrinsically classical in nature. In that case, an underlying theory for gravity will be a totally different theory that is not gravity, just as the nonabelian interaction of quarks and gluons is not nuclear physics.\\

\vspace{.2in}
We therefore take the point of view that gravitation is entirely a classical theory and that general relativity is valid down to the Planck scale. This means that spacetime is continuous as long as we are above the Planck scale. At the Planck length, quantum black holes will appear and they act as a natural cutoff to spacetime. They are created by the very act of observing and determining to the level of accuracy of the Planck length such as by using photons of very short wavelengths and therefore of very high energies. The present goal is to construct various fundamental quantum black holes as elementary particles, using the results in general relativity. This requires the use of several black hole theorems that have been discovered in the last 40 years.\\

\newpage
\noindent
{\bf 3 \hspace{.1in} Black Hole Theorems}\\

\noindent
The theorems for classical black holes are the following:\\  

\noindent
1. Singularity theorem  (1965) [4].\\   
2. Area theorem  (1972) [5].\\    
3. Uniqueness theorem  (1975) [6].\\  
4. Positive energy theorem  (1983) [7].\\  
5. Horizon mass theorem  (2005) [8].\\ 

\noindent
The first four theorems are well known since they have been discussed for many years. The latest one, horizon mass theorem, is only known recently and it is found to be relevant to the construction of quantum black holes.\\
 
   The mass of a black hole depends on where the observer is. The horizon mass theorem states that for all black holes: neutral, charged or rotating, the horizon mass is always {\em twice} the irreducible mass observed at infinity. In notation, it is given as
\begin{equation}
M(r_{+}) = 2M_{irr}.
\end{equation}
Here $M(r_{+})$ is the mass observed at the horizon $r_{+}$. The horizon mass is the mass which cannot escape from the horizon of a neutral, charged or rotating black hole. The irreducible mass $M_{irr}$ is the final mass of a charged or rotating black hole when its charge or angular momentum is removed by adding external particles to the black hole. It is the mass observed at infinity. A surprising consequence of this theorem is that the electrostatic and rotational energies of a general black hole are all external quantities. They are absent inside the black hole. This leads to the remarkable result that a charged black hole does not carry any electric charges inside but they stay at the surface of the black hole. Similarly, a rotating black hole does not actually rotate but it is the external space which is undergoing rotation. A distant observer sees the total energy of the black hole and the energy in the surrounding space and therefore determines an asymptotic mass $M_{\infty}$  that he considers to be the physical mass. This mass becomes the mass $M$  in common notation which appears in the metrics of the Schwarzschild, Reissner-Nordstr\"{o}m and the Kerr black holes.\\

       The horizon mass theorem is crucial for understanding Hawking radiation.  This is because black hole radiation is only possible if the horizon mass is greater than the asymptotic mass since it takes the equivalent of the entire rest mass energy for a neutral particle released near the horizon to reach infinity [9].  Therefore no black hole radiation is possible if the horizon mass is equal to the asymptotic mass. The situation is similar to the photoelectric effect in which the incident photon must have a greater energy than that of the ejected electron in order to overcome binding. Without black hole radiation, the Second Law of Thermodynamics is lost. For quantum black holes, the horizon mass theorem is necessary in order to determine which black holes will be stable and which ones will disintegrate.\\

\vspace{.3in}
\noindent
{\bf 4 \hspace{.1in} Quantum Black Holes}\\

\noindent
Quantum black holes have many characteristics of elementary particles. They have a mass of the Planck mass, a radius of the Planck length, a lifetime of the Planck time or they can be absolutely stable. They may possess a spin which is integer or half-integer. They may also possess an electric charge. As black holes, they have the additional definition of area and intrinsic entropy that ordinary particles do not possess.  We define the fundamental quantum black holes to be the smallest black holes possible such that their radii are exactly equal to the Planck length. We believe that the Planck length is the smallest scale of observation if there is any physical meaning to this quantity.\\
  
Quantum black holes may be created in ultra-high energy collisions or in the Big Bang. Several fundamental types have been constructed with some remarkable properties. They are:
\begin{enumerate}
\item
Planck-charge case - this is a Planck-size black hole carrying maximum electric charge but no spin.  It is absolutely stable and cannot emit any radiation because its horizon mass is equal to its asymptotic mass. The Planck-charge is defined to be $Q_{Pl} = {\sqrt G}M_{Pl} = \sqrt {\hbar c}$ .
\item  Spin-0 case - a Planck-size black hole with no spin and no electric charge. It will disintegrate immediately after it is formed and become Hawking radiation. Its observable signature may be seen from its radiation in the form of an enormous number of particles with very high energies in all directions.
\item  Spin-1/2 case - a Planck-size rotating black hole carrying angular momentum $\hbar/2$ and electric charge ${\sqrt3}Q_{Pl}/2$, and magnetic moment $\mu = {\sqrt3}Q_{Pl}l_{Pl}/4$. It is unstable and it will decay into a burst of elementary particles. A spin-1/2 black hole is by definition a fermion according to a distant observer in an asymptotically flat spacetime.
\item  Spin-1 case - a Planck-size rotating black hole with angular momentum $\hbar$ but no charge. It will also decay into a burst of elementary particles.
\end{enumerate}

\noindent
Microscopic black holes with higher mass and larger size than the fundamental types may be constructed  following the condition
\begin{equation}
\left( \frac{GM}{c} \right)^{2} \geq  G\left( \frac{Q}{c} \right)^{2} + \left( \frac{J}{M} \right)^{2} ,
\end{equation}
where $Q$  is the electric charge and $J$ is the angular momentum. The area of a general Kerr-Newman black hole is given by
\begin{equation}
A = \frac{8\pi G^{2}M^{2}}{c^{4}} \left( 1 + \sqrt{ 1 -  \frac{Q^{2}}{GM^{2}} - \frac{J^{2}c^{2}}{G^{2}M^{4}} } \right )
    - \frac{4 \pi GQ^{2}}{c^{4}} .
\end{equation}
There have been attempts to quantize the area of a black hole in terms of a basic unit of area such as the Planck area
$l_{Pl}^{2}$ and leading to the quantization of the black hole mass. However, these efforts often result in unphysical spins such as transcendental and imaginary numbers that are not found in quantum mechanics. If one should impose the physical integer and half-integer spin condition on the black hole then the result would not achieve the desired quantization of area. At present, black hole quantization remains only a conjecture.\\ 

\vspace{.3in}
\noindent
{\bf 5 \hspace{.1in} Ultra-high Energy Cosmic Rays}\\

\vspace{.2in}
\noindent
In 1966, Greisen [10],  Zatsepin and Kuzmin [11] derived a theoretical upper limit of the energy of cosmic rays by considering the interaction of protons with cosmic microwave background photons. Successive collisions with photons would result in significant energy loss for the protons traveling in intergalactic space and the spectrum of cosmic rays would show a flux suppression above $6 \times 10^{19}$ eV. This is known as the GZK effect.\\
 
In the past decade, several ultra-high energy cosmic rays experiments have been carried out to measure the highest energies of cosmic rays detected in Earth's atmosphere.  The AGASA experiment first reported several events that are above the GZK limit [12].  The HiRes experiment showed the first evidence of the GZK cutoff but reported no correlation with nearby astrophysical sources [13].  The Pierre-Auger Collaboration observed a number of events supporting the GZK cutoff and also found a correlation between these events and nearby active galactic nuclei [14].  The determination of the GZK cutoff in these experiments depends on statistics, i.e. how many events are observed above the limit and how many below the limit based on a particular detection technique in each experiment. A preponderance of events observed below the limit would indicate a cutoff. At this stage, there is hemispherical difference in the nature of the sources and the total number of events observed in all experiments is rather small. The ultimate nature of these cosmic ray events is still to be established. It is very likely that the cutoff is real. However, if there are events that persist above the GZK limit and which cannot be explained when there are no near-Earth astrophysical sources, then quantum black holes in the neighborhood of the Galaxy could resolve this paradox. In particular, those Planck-charge quantum black holes that are absolutely stable can annihilate with opposite ones to produce powerful bursts of elementary particles with very high energies. This is a possibility worth considering.\\

\vspace{.2in}
Quantum black holes could also play a role as dark matter in cosmology without having to resort to new interactions and exotic particles. In this case, a pair of similar Planck-charge quantum black holes shows the remarkable property that their electrostatic repulsion exactly cancels their gravitational attraction so that there is no effective potential between them at any distance. A collection of these quantum black holes in a finite volume behaves like a non-interacting gas. Each constituent has exactly one unit of Planck mass and one unit of Planck length as radius. They are the closest form of matter to the ideal `point particles' invented in classical mechanics. Quantum black holes in our construction could have a real existence. Their potential discovery would lead us to a deeper understanding of the ultimate nature of spacetime and matter.\\


\begin{thebibliography}{99}
\bibitem{1} G. 't Hooft, {\em Nucl. Phys.} {\bf B256}, 727 (1985).
\bibitem{2} T. Rothman and S. Boughn, {\em Found. Phys.} {\bf 36}, 1801 (2006).
\bibitem{3} Y.K. Ha, {\em arXiv}: 0812.5012.
\bibitem{4} R. Penrose, {\em Phys. Rev. Lett.} {\bf 14}, 57 (1965);
            S.W. Hawking, {\em Phys. Rev. Lett.} {\bf 15}, 689 (1965).
\bibitem{5} S.W. Hawking, {\em Commun. Math. Phys.} {\bf 25}, 152 (1972).
\bibitem{6} D.C. Robinson, {\em Phys. Rev. Lett.} {\bf 34}, 905 (1975).
\bibitem{7} R. Schoen and S.T. Yau, {\em Commun. Math. Phys.} {\bf 80}, 381 (1981); 
            G.W. Gibbons, S.W. Hawking, G.T. Horowitz and M.J. Perry, {\em Commun. Math. Phys.} {\bf 88}, 295 (1983).
\bibitem{8} Y.K. Ha, {\em Int. J. Mod. Phys. D} {\bf 14}, 2219 (2005); 
            Y.K. Ha, {\em arXiv}: gr-qc/0703130.
\bibitem{9} Y.K. Ha, {\em Gen. Rel. Grav.} {\bf 35}, 2045 (2003).
\bibitem{10} K. Greisen, {\em Phys. Rev. Lett.} {\bf 16}, 748 (1966).
\bibitem{11} G.T. Zatsepin and V.A. Kuzmin, {\em JEPT Lett}. {\bf 4}, 78 (1966).
\bibitem{12} M. Takeda et al., {\em Astropart. Phys.} {\bf 19}, 447 (2003).
\bibitem{13} The HiRes Collaboration, {\em Phys. Rev. Lett.} {\bf 100}, 101101 (2008).
\bibitem{14} The Pierre Auger Collaboration, {\em Science} {\bf 318}, 938 (2007).
\end{thebibliography}
\end{document}